\documentclass[journal,10pt]{IEEEtran}

\usepackage[acronym,toc,shortcuts]{glossaries}
\usepackage[inline]{enumitem}
\usepackage{multirow}
\usepackage{amssymb}
\usepackage{multirow}
\usepackage{array}
\usepackage{caption}
\usepackage{lipsum}
\usepackage{mathtools}
\usepackage{cuted}
\usepackage{breqn}
\usepackage{graphicx}
\usepackage[outdir=./]{epstopdf}
\usepackage[english]{babel}
\usepackage[autostyle]{csquotes}
\usepackage{amsmath}
\usepackage{verbatim}
\usepackage[graphicx]{realboxes}
\usepackage{breqn}
\usepackage{subfig}
\usepackage{euscript}
\usepackage{color}
\usepackage[lined,linesnumbered,noend,ruled]{algorithm2e}
\usepackage{url}
\usepackage{cite}
\usepackage{adjustbox}						
\usepackage{booktabs,array}	
\usepackage{comment}
\usepackage{multicol,amsmath,lipsum}
\usepackage[margin=2cm]{geometry}
\usepackage{resizegather}

\makeglossaries
\newacronym{BER}{BER}{bit-error-rate}
\newacronym{APD}{APD}{avalanche photo detector}
\newacronym{AWG}{AWG}{arbitrary signal generator}
\newacronym{CC}{CC}{Convolutional Code}
\newacronym{DC}{DC}{direct current}
\newacronym{CamCom}{CamCom}{optical camera communications}
\newacronym{CIR}{CIR}{channel impulse response}
\newacronym{CLT}{CLT}{Central Limit Theorem}
\newacronym{CMOS}{CMOS}{Complementary Metal Oxide Semiconductor}
\newacronym{COTS}{COTS}{commercial off-the-shelf}
\newacronym{C-V2X}{C-V2X}{cellular vehicle to everything}
\newacronym{CSK}{CSK}{Color Shift Keying}
\newacronym{DSRC}{DSRC}{Dedicated Short Range Communication}
\newacronym{DMT}{DMT}{discrete-multi-tone}
\newacronym{DRL}{DRL}{day time running light}
\newacronym{DPD}{DPD}{digital pre-distorter}
\newacronym{FEC}{FEC}{Forward Error Correction}
\newacronym{FLP}{FLP}{fast locking pattern}
\newacronym{FoV}{FoV}{field of view}
\newacronym{FPGA}{FPGA}{Field Programmable Gate Array}
\newacronym{FWHM}{FWHM}{full width at half maximum}
\newacronym{fps}{fps}{frame per second}
\newacronym{IF}{IF}{intermediate frequency}
\newacronym{IM/DD}{IM/DD}{intensity modulation and direct detection}
\newacronym{ISI}{ISI}{intersymbol interference}
\newacronym{ITS}{ITS}{intelligent transportation systems}
\newacronym{KNN}{KNN}{k-nearest neighbor classifier}
\newacronym{LED}{LED}{light emitting diode}
\newacronym{LNA}{LNA}{low noise amplifier}
\newacronym{LoS}{LoS}{line-of-sight}
\newacronym{LUT}{LUT}{look-up table}
\newacronym{LTE}{LTE}{Long Term Evolution}
\newacronym{MAC}{MAC}{medium access control}
\newacronym{ML}{ML}{machine learning}
\newacronym{MLP}{MLP}{multilayer perceptron}
\newacronym{MP}{MP}{Memory Polynomial}
\newacronym{MPC}{MPC}{multi path component}
\newacronym{MCS}{MCS}{modulation coding schemes}
\newacronym{MIMO}{MIMO}{multiple input multiple output}
\newacronym{MSE}{MSE}{mean square error}
\newacronym{NLoS}{NLoS}{non-line-of-sight}
\newacronym{NOMA}{NOMA}{non-orthogonal multiple access}
\newacronym{NN}{NNet}{neural network}
\newacronym{OCC}{OCC}{optical  camera  communications}
\newacronym{OLED}{OLED}{organic light emitting diodes}
\newacronym{OOK}{OOK}{on-off keying modulation}
\newacronym{OWC}{OWC}{Optical Wireless Communication}
\newacronym{QAM}{QAM}{quadrature amplitude modulation}
\newacronym{PAM}{PAM}{pulse amplitude modulation}
\newacronym{PD}{PD}{photodetector}
\newacronym{PHR}{PHR}{physical header}
\newacronym{PHY}{PHY}{physical layer}
\newacronym{PMT}{PMT}{photomultiplier tube}
\newacronym{PPM}{PPM}{pulse position modulation}
\newacronym{PWM}{PWM}{pulse width modulation}
\newacronym{PSDU}{PSDU}{physical service data unit}
\newacronym{RBF}{RBF}{radial basis function}
\newacronym{RF}{RF}{radio frequency}
\newacronym{RLL}{RLL}{Run-Length Limited}
\newacronym{RMSE}{RMSE}{root mean square error}
\newacronym{RS}{RS}{Reed Solomon}
\newacronym{RSS}{RSS}{received signal strength}
\newacronym{RU}{RU}{Receiver unit}
\newacronym{SDR}{SDR}{Software-Defined Radio}
\newacronym{SHR}{SHR}{synchronization header}
\newacronym{SNR}{SNR}{signal-to-noise ratio}
\newacronym{SPAD}{SPAD}{single photon avalanche diode}
\newacronym{TDP}{TDP}{topology dependent pattern}
\newacronym{TU}{TU}{Transmitter Unit}
\newacronym{USRP}{USRP}{Universal Software Radio Peripheral}
\newacronym{V-I}{V-I}{voltage-current}
\newacronym{V2I}{V2I}{vehicle to infrastructure}
\newacronym{V2V}{V2V}{vehicle to vehicle}
\newacronym{V2X}{V2X}{vehicle to everything communications}
\newacronym{VLC}{VLC}{visible light communications}
\newacronym{V2P}{V2P}{vehicle to pedestrian}
\newacronym{VLP}{VLP}{visible light positioning}
\newacronym{VNA}{VNA}{vector network analyzer}
\newacronym{VPPM}{VPPM}{variable pulse position modulation}
\newacronym{WDGF}{WDGF}{weighted double gamma function}
\newacronym{DCO-OFDM}{DCO-OFDM}{direct current biased optical orthogonal frequency division multiplexing}
\newacronym{OFDM}{OFDM}{orthogonal frequency division multiplexing}
\newacronym{PAPR}{PAPR}{peak-to-average-power ratio}
\newacronym{PA}{PA}{power amplifier}
\newacronym{MAPE}{MAPE}{mean absolute percentage error}
\newacronym{AWGN}{AWGN}{additive white Gaussian noise}
\newacronym{FOV}{FOV}{field-of-view}
\newacronym{PSD}{PSD}{power spectral density}
\newacronym{IFFT}{IFFT}{inverse fast Fourier transform}
\newacronym{FFT}{FFT}{fast Fourier transform}
\newacronym{VVLC}{V-VLC}{vehicular visible light communications}

\begin{document}
\title{Measurement Based Non-Line-Of-Sight \\ Vehicular Visible Light Communication  Channel Characterization}

\author{Bugra~Turan, Omer~Narmanlioglu, Osman~Nuri~Koc, Emrah~Kar,~\IEEEmembership{Student Member,~IEEE}, {Sinem~Coleri,~\IEEEmembership{Senior~Member,~IEEE}}, and Murat~Uysal,~\IEEEmembership{Fellow,~IEEE}

\thanks{B.Turan, O. N. Koc and E. Kar are with the Koc University Ford Otosan Automotive Technologies Laboratory, Sariyer, Istanbul, 34450, Turkey. \protect\\
E-mail: bturan14@ku.edu.tr, okoc@ku.edu.tr, ekar@ku.edu.tr}
\thanks{S. Coleri is with the Department of Electrical and Electronics Engineering, Koc University, Sariyer, Istanbul 34450, Turkey. \protect\\
E-mail: scoleri@ku.edu.tr}
\thanks{O. Narmanlioglu and M. Uysal are with the Department of Electrical and Electronics Engineering, Ozyegin University, Istanbul 34794, Turkey. \protect\\ 
E-mail: omernarmanlioglu@gmail.com, murat.uysal@ozyegin.edu.tr}% <-this % stops a space

\thanks{This work was supported by CHIST-ERA grant CHIST-ERA-18-SDCDN-001, the Scientific and Technological Council of Turkey 119E350 and Ford Otosan.}
% <-this % stops a space
}
%\markboth{Journal of \LaTeX\ Class Files,~Vol.~XX, No.~XX, September~2017}%

\maketitle

\begin{abstract}

% \Ac{VVLC} aims to provide secure complementary \ac{V2X} to increase road safety and traffic efficiency. \ac{VVLC} provides directional transmissions, mainly enabling \ac{LoS} communications. However, reflections due to nearby objects enable \ac{NLoS} transmissions, extending the usage scenarios beyond \ac{LoS}. In this paper, we propose wide-band measurement based \ac{NLoS} channel characterization, and evaluate the performance of \ac{DCO-OFDM} \ac{VVLC} scheme for \ac{NLoS} channel. We propose a distance based \ac{NLoS} \ac{VVLC} channel path loss model considering reflection surface characteristics and \ac{NLoS} \ac{VVLC} \ac{CIR} incorporating the temporal broadening effect due to vehicle reflections through \ac{WDGF}. The proposed path loss model yields higher accuracy up to 14 dB when compared to single order reflection model whereas \ac{CIR} model estimates the \ac{FWHM} up to 2 ns accuracy. We further demonstrate that the target BER of $10^{-3}$ can be achieved up to 7.86 m, 9.79 m and 17.62 m distances for black, orange and white vehicle reflection induced measured \ac{NLoS} \ac{VVLC} channels for \ac{DCO-OFDM} transmissions. 

Vehicular visible light communication (V-VLC) aims to provide secure complementary \ac{V2X} to increase road safety and traffic efficiency. V-VLC provides directional transmissions, mainly enabling line-of-sight (LoS) communications. However, reflections due to nearby objects enable non-line-of-sight (NLoS) transmissions, extending the usage scenarios beyond LoS. In this paper, we propose wide-band measurement based NLoS channel characterization, and evaluate the performance of direct current biased optical orthogonal frequency division multiplexing (DCO-OFDM) V-VLC scheme for NLoS channel. We propose a distance based NLoS V-VLC channel path loss model considering reflection surface characteristics and NLoS V-VLC channel impulse response (CIR) incorporating the temporal broadening effect due to vehicle reflections through weighted double gamma function. The proposed path loss model yields higher accuracy up to 14~dB when compared to single order reflection model whereas CIR model estimates the full width at half maximum up to 2~ns accuracy. We further demonstrate that the target bit-error-rate of $10^{-3}$ can be achieved up to 7.86~m, 9.79~m and 17.62~m distances for black, orange and white vehicle reflection induced measured NLoS V-VLC channels for DCO-OFDM transmissions.

\end{abstract}


\IEEEpeerreviewmaketitle

\section{Introduction}

Vehicular communications enable exchange of traffic, road and vehicle information between vehicles, infrastructure and other road users through \ac{V2V}, \ac{V2I} and \ac{V2P} communications. Currently, \ac{RF} based \ac{C-V2X}~\cite{molina2017lte} and IEEE 802.11p~\cite{arena2020review} technologies are standardized to support \ac{V2X} applications. On the other hand, omnipresence of \glspl{LED} in the vehicle lighting paves the way for \ac{VVLC} as a complementary technology to the existing schemes. Since \ac{VVLC} provides \ac{RF} interference, jam and spoof free directional optical communications, it is foreseen to be a strong candidate to enable secure vehicular communications~\cite{memedi2020vehicular}.

\ac{VVLC} is mainly considered to be a \ac{LoS} technology due to its directional propagation characteristics. However, \ac{NLoS} \ac{VVLC} transmissions are also demonstrated to increase \ac{RSS} of \ac{LoS} \ac{VVLC} link through object reflections~\cite{tebruegge2019empirical}. Therefore, \ac{NLoS} \ac{VVLC} can be considered to support close proximity \ac{V2V} safety applications such as blind spot warning, lane change warning and intention sharing by taking advantage of nearby vehicle reflected optical signals.  

\ac{VVLC} system performance mainly depends on the \ac{VVLC} propagation channel characteristics, which distort the optical signals through attenuation, scattering and reflection. Thus, \ac{VVLC} channel modeling is extensively studied in the literature, based on the \ac{LoS} transmission assumptions \cite{karbalayghareh2020channel, eldeeb2021vehicular,al2021impact}. \cite{karbalayghareh2020channel} proposes a ray tracing simulation based \ac{LoS} path loss model incorporating headlight radiation pattern, \cite{eldeeb2021vehicular} presents taillight measurement based path loss model and \cite{al2021impact} provides a statistical path loss distribution model with the consideration of asymmetric headlight radiation pattern. On the other hand, only a few studies investigate the \ac{NLoS} \ac{VVLC} channel  \cite{tebruegge2019empirical,luo2015performance}, where \cite{tebruegge2019empirical} demonstrates the incremental effect of highly reflective road surface on \ac{VVLC} \ac{RSS} and \cite{luo2015performance} shows the \ac{BER} performance degradation of \ac{VVLC} due to road surface specular reflections. However, the current literature lacks the characterization of \ac{NLoS} \ac{VVLC} channel due to vehicle reflections, where the vehicle surface mainly determines the path loss and signal distortions. On the other hand, \ac{CIR} temporal broadening due to \ac{NLoS} scattering, which determines the achievable data rate and bandwidth of the \ac{VVLC} systems has not yet been investigated for \ac{VVLC} channels. Moreover, empirical channel models based on the low bandwidth optical power measurements by optical power meters \cite{eldeeb2021vehicular} and \ac{RSS} measurements at single tone frequencies \cite{tebruegge2019empirical} fail to address the channel behavior for a wide range of \ac{LED} modulation frequencies. 

In this paper, we propose wide-band frequency sweep measurement based \ac{CIR} and path loss characterization for \ac{NLoS} \ac{VVLC} channels due to vehicle body reflections. Moreover, we investigate the \ac{BER} performance of \ac{DCO-OFDM} based \ac{VVLC} system for \ac{NLoS} channels for different modulation order and background noise levels. We demonstrate that the proposed \ac{NLoS} \ac{VVLC} channel path loss model provides higher accuracy than the widely considered single order reflection based \ac{NLoS} channel model \cite{komine2004fundamental} for indoor \ac{VLC}, since \ac{NLoS} channel model for \ac{VVLC} has not yet been proposed in the literature. The original contributions of this paper are threefold: \begin{itemize}
    \item We provide a distance based \ac{NLoS} \ac{VVLC} channel path loss model with the consideration of pure vehicle surface   reflections, for the first time in the literature.
    \item We characterize \ac{CIR} of the \ac{NLoS} \ac{VVLC} through \ac{WDGF} by using measurement data from different vehicle reflection induced channels. Thus, for the first time in the literature, we investigate the temporal broadening effect of \ac{VVLC} \ac{CIR} due to vehicle reflections.
    \item We provide attainable \ac{NLoS} \ac{VVLC} distance for the target \ac{BER} performance of \ac{DCO-OFDM} scheme under the consideration of measured channel gains for a wide range of \ac{LED} modulation frequencies.
\end{itemize}  

The rest of the paper is organized as follows. Section~\ref{chmodel} describes the considered \ac{DCO-OFDM} transmission system model. Section~\ref{sec:measmodel} provides the methodology for measurement based \ac{NLoS} channel modeling. Section~\ref{sec:performance} presents the proposed path model and \ac{WDGF} based \ac{NLoS} \ac{CIR} characterization. Section~\ref{sec:dcoofdm} presents simulation results for \ac{DCO-OFDM} \ac{VVLC} for the empirical \ac{NLoS} \ac{VVLC} channel. Finally, Section~\ref{sec:conc} concludes the paper. 

\noindent
%\textit{{Notations:}} $[.]^{\rm{T}}$ and $(.)^*$ denote transpose and complex conjugate operations, respectively, and $\otimes$ denotes linear convolution.

\section{System Model}\label{chmodel}

%In this section, widely used single order reflection based \ac{NLoS} \ac{VLC} channel path loss model and the \ac{DCO-OFDM} based transmission model are provided. The single order reflection based path loss model is considered as a benchmark to be compared with the proposed model, whereas the performance of \ac{DCO-OFDM} based \ac{VVLC} system is evaluated for the proposed \ac{NLoS} channels. %path loss expression obtained by measurement data least-squares fitting is provided. Moreover, distance and reflection surface dependent \ac{CIR} of \ac{NLoS} \ac{VVLC} is characterized.
In this section, \ac{DCO-OFDM} based transmission model is provided, where the performance of \ac{DCO-OFDM} based \ac{VVLC} system is evaluated for the proposed \ac{NLoS} channels.

%\paragraph{DCO-OFDM Transmissions System Model}

In \ac{DCO-OFDM}, binary information is first mapped to complex symbols $s_1 \textrm{ } s_2 \textrm{ } ... \textrm{ } s_{N/2-1}$ based on the deployed constellation scheme, such as $M-$order quadrature amplitude modulation (QAM), where $N$ is the number of subcarriers. To ensure that the output of \ac{IFFT} is real-valued, Hermitian symmetry is imposed on the transmitted signal as, 
\begin{equation}
    {{\bf{X}}_{\bf{S}}} = {\left[ { 0 \textrm{ } {s_1} \textrm{ } {s_2}  \textrm{ } ... \textrm{ } {s_{N/2 - 1}} \textrm{ } 0 \textrm{ }  s_{N/2 - 1}^* \textrm{ } ... \textrm{ }  s_2^* \textrm{ }  s_1^* } \right]^{\rm{T}}}, \label{hermitianSym}
\end{equation}
\noindent
where $(.)^{*}$ and $[.]^{\rm T}$ denote complex conjugate and transpose operations, respectively. After $N-$point \ac{IFFT} operation where the output is denoted by $x_S[n]$, a cyclic prefix with length $N_{\mathrm{CP}}$, which is greater than or equal to the delay spread of the channel, is appended to the beginning of each \ac{DCO-OFDM} frame and parallel streams are converted to serial. The resulting \ac{DCO-OFDM} signal is 
\begin{equation}
x_S(t) = \sum_{n=0}^{N+N_{\mathrm{CP}}-1}x_S[n] \delta(t-nT_{s}),
\end{equation}
where $T_{s}$ is the sampling interval and $x_S(t)$ has the average electrical transmit signal power of $P_{\rm T}$. The modulated signal is \ac{DC} biased and transmitted via \ac{LED}. 
The time domain received signal at the \ac{PD} is
\begin{equation}
y(t) = R  x_S(t)  \otimes {h}(t) + v_n(t), \label{tr_model}
\end{equation}
where $R$ is the responsivity of \ac{PD} (A/W), $h(t)$ is the optical channel response obtained by the measurements, and $v_n(t)$ is \ac{AWGN} with zero mean and $\sigma _N^2=N_0W$ variance, where $N_0$ is noise \ac{PSD} and $W$ denotes the system bandwidth. In~(\ref{tr_model}), $\otimes$ denotes the convolution operation.

At the receiver, the captured optical signals are converted to parallel streams and cyclic prefix is removed. Then, the $N-$point \ac{FFT} is applied on the output. %output signal on the $k^{\rm th}$ subcarrier of (\ref{tr_model}).

\section{Measurement Based \ac{NLoS} Channel Modeling}\label{sec:measmodel}

Vehicle surfaces are made up of different materials and follow various geometries, where \ac{VVLC} signal reflection characteristics vary between vehicles. To incorporate the vehicle surface dependent reflection characteristics, a measurement based approach is employed with the following steps: \begin{enumerate*}[label=(\roman*)]
\item characterize the \ac{LED} modulation bandwidth and ambient noise to determine measurement parameters of maximum sweep frequency and intermediate frequency bandwidth $({\rm IF \; BW})$, \item measure the \ac{LoS} channel frequency response to obtain channel gain values as a benchmark to evaluate \ac{NLoS} channel, \item measure \ac{NLoS} channel gains for path loss and \ac{CIR} characterization, 
\item post-process measurement data to obtain \ac{NLoS} channel path loss and~\ac{CIR}.
\end{enumerate*}

The \ac{LED} modulation bandwidth determines the maximum measurable frequency of the channel, whereas \ac{VNA} intermediate frequency bandwidth selection determines the noise floor of the measurements. The \ac{LoS} channel gain ($S_{21}$) values constitute reflection free, non-distorted channel frequency response measurements, which are used to evaluate reflection induced \ac{NLoS} measurements. The \ac{NLoS} $S_{21}$ measurements are post processed to extract distance and reflection surface dependent path loss with \ac{CIR} temporal characteristics. 

\begin{figure}[h]
\centering
\includegraphics[clip,trim=0 0 0 0,width=0.8\linewidth]{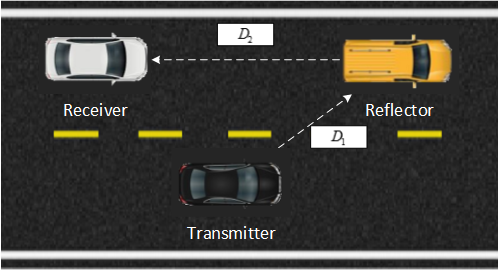}
\caption{NLoS Scenario}
\label{Fig:scenario}
\end{figure}

To capture \ac{NLoS} components from vehicle reflections, a  vehicle \ac{LED} \ac{DRL} in the nearby lane is directed towards the rear of the reflector vehicle. The receiver vehicle is positioned behind the reflector vehicle in the same lane, where \ac{LoS} link does not exist between the transmitter and receiver vehicle. The transmitter-reflector vehicle distance ($D_{1}$) is fixed, whereas the distance between reflector and receiver vehicle ($D_{2}$) is varied. Three different reflection surfaces, orange light commercial vehicle (Ford Courier 2019), white (BMW 320i 2014) and black passenger vehicle (Fiat Tipo Station Wagon 2018) bodies are considered as reflection surfaces to model the \ac{NLoS} channel. The measurement scenario is depicted in Fig.~\ref{Fig:scenario}.

The \ac{LoS} channel path loss of the measurement setup is obtained with the limitation of \ac{PD}'s \ac{FoV} with a tube similar to the work in \cite{tebruegge2019empirical}, at the same transmitter-receiver height. Then, \ac{NLoS} channel measurements are conducted to obtain \ac{LED} modulation frequency dependent channel gains. The path loss and \ac{CIR} are  calculated at the post-processing stage. To minimize random variations of the received signal due to measurement equipment, all measured frequency responses are averaged over 10 realizations per measurement location.

At the transmitter side, the sounding signals are fed into the driver block from Port 1 of the \ac{VNA}. The driver block amplifies the sounding signals and adds \ac{DC} bias before feeding to the \ac{LED}. At the receiver side, \ac{PD} captures the signals and feed into the Port 2 of the \ac{VNA}, yielding frequency dependent $S_{21}$ values. 

\paragraph{Path Loss Characterization} The path loss is derived as the ratio of the received electrical power to the transmitted electrical power over the measurement bandwidth. The path loss from the measurement data is obtained by~\cite{martinez2016channel} 
\begin{equation}
    {\rm PL \; [dB]} = -10{\rm log}_{10}{\left(\frac{1}{N_{\rm f}}\sum_{n=0}^{N_{\rm f}-1} {|S_{\rm 21}(f_{\rm min}+n\Delta f)|^2}\right)},
\end{equation}
 
\noindent where $N_{\rm f}$ is the total number of $S_{\rm 21}$ values from minimum measurement frequency, $f_{\rm min}$, to maximum measurement frequency, $f_{\rm max}$, due to \ac{LED} modulation bandwidth limitations, and $\Delta f$ is the step frequency.

The reflection coefficient ($\rho$) of the considered surfaces is calculated by the ratio of the measured \ac{NLoS} reflected surface channel gain to pure \ac{LoS} channel gain at the reference distance of 2 m. The reflection coefficient demonstrates the reflectivity of the surface, where higher values indicate more reflective surfaces (i.e., brighter color).

\begin{comment}
\begin{table}[]
\centering
\caption{Reflection Surface Coefficient}
\begin{tabular}{|l||l|l|}
\hline
\textbf{Reflection Surface}  & \textbf{Coefficient} \\ \hline
White Wall & 0.0936 \\ \hline
%Whiteboard & 0.1424        \\ \hline
Vehicle Mirror &    0.2134 \\ \hline
%Wheel & 0.0148        \\ \hline
%Fmax headlight & 0.0270         \\ \hline
Orange Light Commercial Vehicle Body & 0.0279         \\ \hline
Black Passenger Car Body & 0.0066         \\ \hline
\end{tabular}
\label{tab:PLE}
\end{table}
\end{comment}

\begin{table}[]
\centering
\caption{\ac{NLoS} path loss model and reflection coefficients}
\begin{tabular}{|l|c|c|c|c|}
\hline
\multirow{2}{*}{\textbf{Surface}} & \multicolumn{4}{c|}{\textbf{Parameters}}                                                                                                                                                                        \\ \cline{2-5} 
                                  & \multicolumn{1}{c|}{\textbf{$\alpha$}} & \multicolumn{1}{c|}{\textbf{$\beta$}} & \multicolumn{1}{c|}{\textbf{n}} & \multicolumn{1}{l|}{\textbf{\begin{tabular}[c]{@{}l@{}}Reflection\\ Coefficient\end{tabular}}} \\ \hline
White Car                             & 0.9185                              & 4.703                              & 0.7189                              & 0.0774                                                                                            \\ \hline
Orange Car                           & 0.7871                              & 5.477                              & 0.9998                              &   0.0243                                                                                      \\ \hline
Black Car                            & 0.7516                              & 5.384                              & 0.9238                              & 0.0156                                                                                           \\ \hline
\end{tabular}
\label{tbl:coef}
\end{table}

\paragraph{\ac{NLoS} \ac{VVLC} \ac{CIR} Characterization}
\label{CIR}

The \ac{NLoS} \ac{CIR} is characterized to evaluate the temporal distortion effects of reflections on \ac{VVLC} signals. To obtain \glspl{CIR} of the channels under consideration, non-measured frequency interval of 0-200 kHz  due to \ac{VNA} limitations, are zero padded and Hermitian symmetry is imposed on the $N$ measured $S_{21}$ parameters. % as 

%\begin{equation}
 %   {{\bf{H}}_{\rm f}} = {\left[ { 0 \textrm{ } {s_{21,1}} \textrm{ } {s_{21,2}} \textrm{ }  ... \textrm{ } {s_{N - 1}} \textrm{ } 0 \textrm{ }  s_{N - 1}^* \textrm{ } ...\textrm{ }  s_{21,2}^* \textrm{ }  s_{21,1}^* \textrm{ } } \right]^{\rm{T}}}. 
%\end{equation} 

The \ac{IFFT} of the zero padded and Hermitian symmetry imposed samples yields \ac{CIR} of the \ac{NLoS} channel. To increase the time resolution of the \ac{CIR}, frequency domain measurement data are zero padded for the frequencies between $f_{\rm max}$, and the target frequency ($f_{\rm target}$), the bandwidth is given by ${\rm BW}=f_{\rm target}-f_{\rm min}$ and time resolution is given by  $t_{\rm res}=1/{\rm BW}$.

The measurement parameters are selected as $f_{\rm min} = 200$~kHz, $f_{\rm max}= 3$~MHz, $\Delta f=700$~Hz, ${\rm IF \; BW}=$ $10$~Hz, noise floor $=-110$~dBm, whereas $t_{\rm res}=1$~ns is selected with the padding of measurement samples. %The channel measurement parameters are provided in Table~\ref{tbl:meas}. 

%\begin{table}[]
%\centering
%\caption{Channel Measurement Setup Specifications}
%\label{tbl:meas}
%\begin{tabular}{lc}
%\hline 
%\textbf{Parameter} & \textbf{Value} \\ \hline
%\multicolumn{2}{c}{\textbf{Transmitter}} \\ \hline
%Headlight 3-dB Bandwidth & 2 MHz \\ 
%DC Bias Voltage & 12 $V$ \\
%Driver Block Input Signal Amplitude & 63 m$V_{pp}$ \\
%Driver Block Total Gain  & 47 dB  \\
%Driver Block Output Signal Amplitude& 14.1 $V_{pp}$ \\
%LED Input Signal Amplitude & 5.6 $ V_{pp}$ \\ 
%LED Optical Transmitted Power & -12 dBm \\ 
%Transmitter Height & 0.7 m \\ \hline
%\multicolumn{2}{c}{\textbf{Receiver}}\\ \hline
%Photodiode & Thorlabs PDA36\\
%Sweep Frequency ($f_{min}$ - $f_{max}$) & 200 $kHz$ - 3 $MHz$ \\
%$\Delta f$  & 700 $Hz$ \\ 
%Number of points & 4001 \\
%Time resolution & 1 $ns$\\
%\Ac{IF} Bandwidth & 100 $Hz$ \\
%Noise Floor & -120 $dBm$\\
%Receiver Height & 0.7 m\\ \hline \hline
%\end{tabular}
%\end{table}

%\textcolor{red}{https://tsapps.nist.gov/publication/get_pdf.cfm?pub_id=927444 3.5 i ekleyelim}

\section{Empirical \ac{CIR} and Path Loss Model}
\label{sec:performance}

In this section, we provide \ac{CIR} tail characterization together with the least-squares fitting based channel path loss model for \ac{NLoS} \ac{VVLC} channel.

\subsection{Channel Impulse Response}
\begin{figure}[h]
\centering
\includegraphics[clip,trim=0 0 0 0,width=\linewidth]{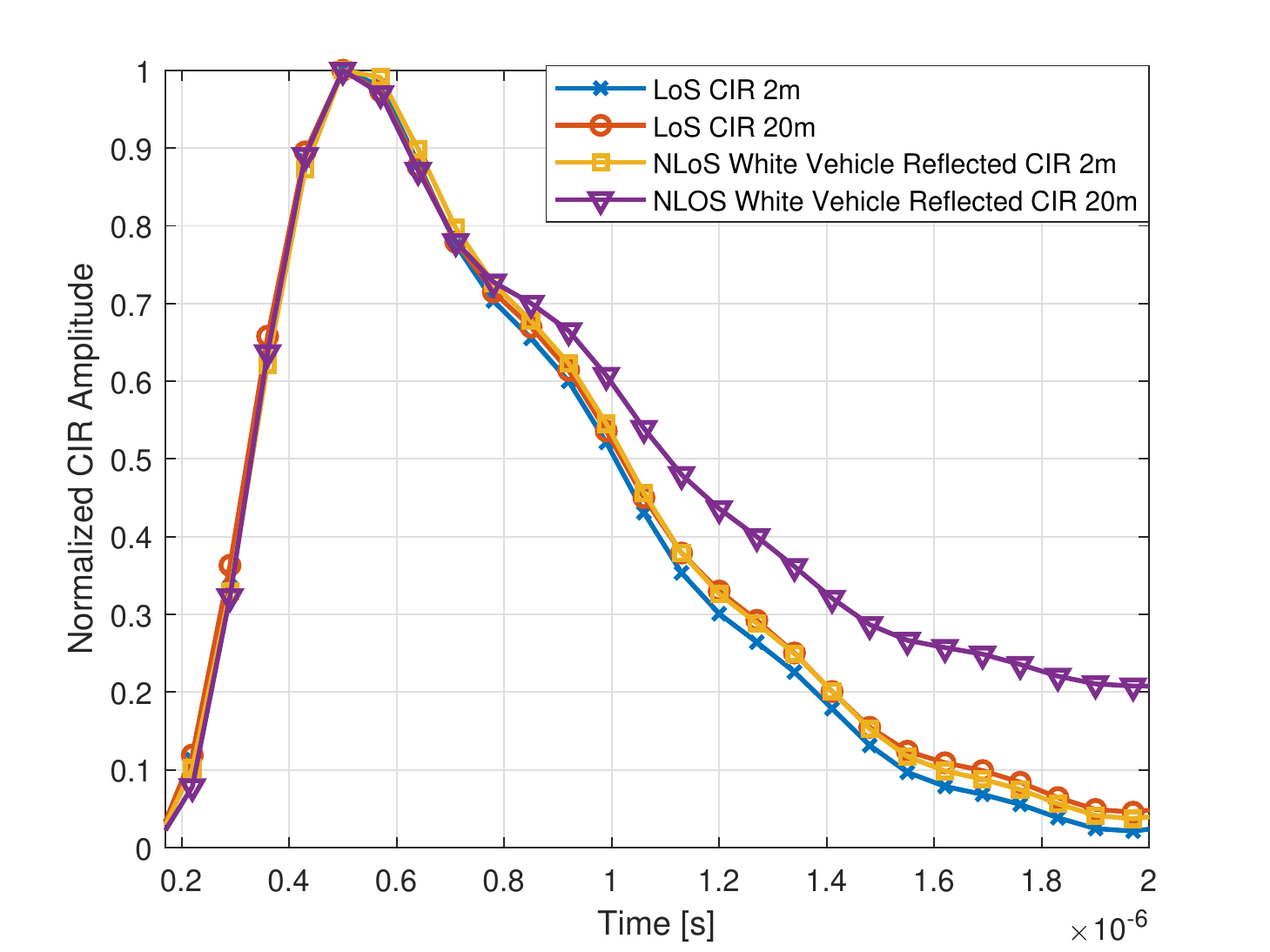}
\caption{LoS and NLoS \ac{CIR} distance dependent temporal broadening.}
\label{Fig:CIRAmplitude}
\end{figure}

The increasing transmitter-receiver for \ac{LoS} and reflector-receiver distances for \ac{NLoS} channels result in longer \ac{CIR} tails due to increased photons scattering (See Fig.~\ref{Fig:CIRAmplitude}). %Since \ac{CIR} broadening determines the achievable data rate and bandwidth of the \ac{VVLC} systems, 
We characterize \ac{CIR} temporal broadening as a function of transmitter-receiver distance for \ac{LoS} and reflector-receiver distance for \ac{NLoS} channels. We adopt \ac{FWHM} metric to quantify \ac{CIR} broadening~\cite{chen2010experimental} and \ac{NLoS} \ac{VVLC} \ac{CIR} is approximated by \ac{WDGF} as~\cite{dong2014impulse}
\begin{equation}
    h(t)=C_{1}\Delta t^{\alpha}{\rm exp}(-C_{2}\Delta t)+C_{3}\Delta t^{\beta}{\rm exp}(-C_{4}\Delta t),
\end{equation}

\noindent where $C_{1}$, $C_{2}$, $C_{3}$, $C_{4}$, $\alpha$, and $\beta$ are model coefficients to be found by least-squares fitting to the measurement data. $\Delta t$ denotes the time scale starting from the \ac{CIR} peak, since the tail determines the \ac{CIR} broadening.

%Fig.\ref{Fig:CIRAmplitude} depicts the normalized \ac{NLoS} \ac{CIR} for the reflector-receiver distance between 2 m to 20 m, where the delay spread increases with the increasing distance from white vehicle reflector surface. 

Fig.~\ref{Fig:CIRAmplitude} demonstrates the \ac{CIR} temporal broadening of \ac{LoS} and \ac{NLoS} channels with the increasing distance. For \ac{LoS} transmissions at 2 m and 20 m transmitter-receiver distances, 671 ns and 690 ns \ac{FWHM} is measured, respectively. On the other hand, white vehicle reflected \ac{NLoS} transmissions yield 696 ns and 778 ns \ac{FWHM} at 2 m and 20 m distances, respecively. Thus, \ac{CIR} broadening due to \ac{NLoS} transmissions can be considered substantial when compared to \ac{LoS} transmissions since delay spread at 2 m \ac{NLoS} channel is almost equivalent to 20 m \ac{LoS} channel.

\begin{figure}[h]
\centering
\includegraphics[clip,trim=0 0 0 0,width=\linewidth]{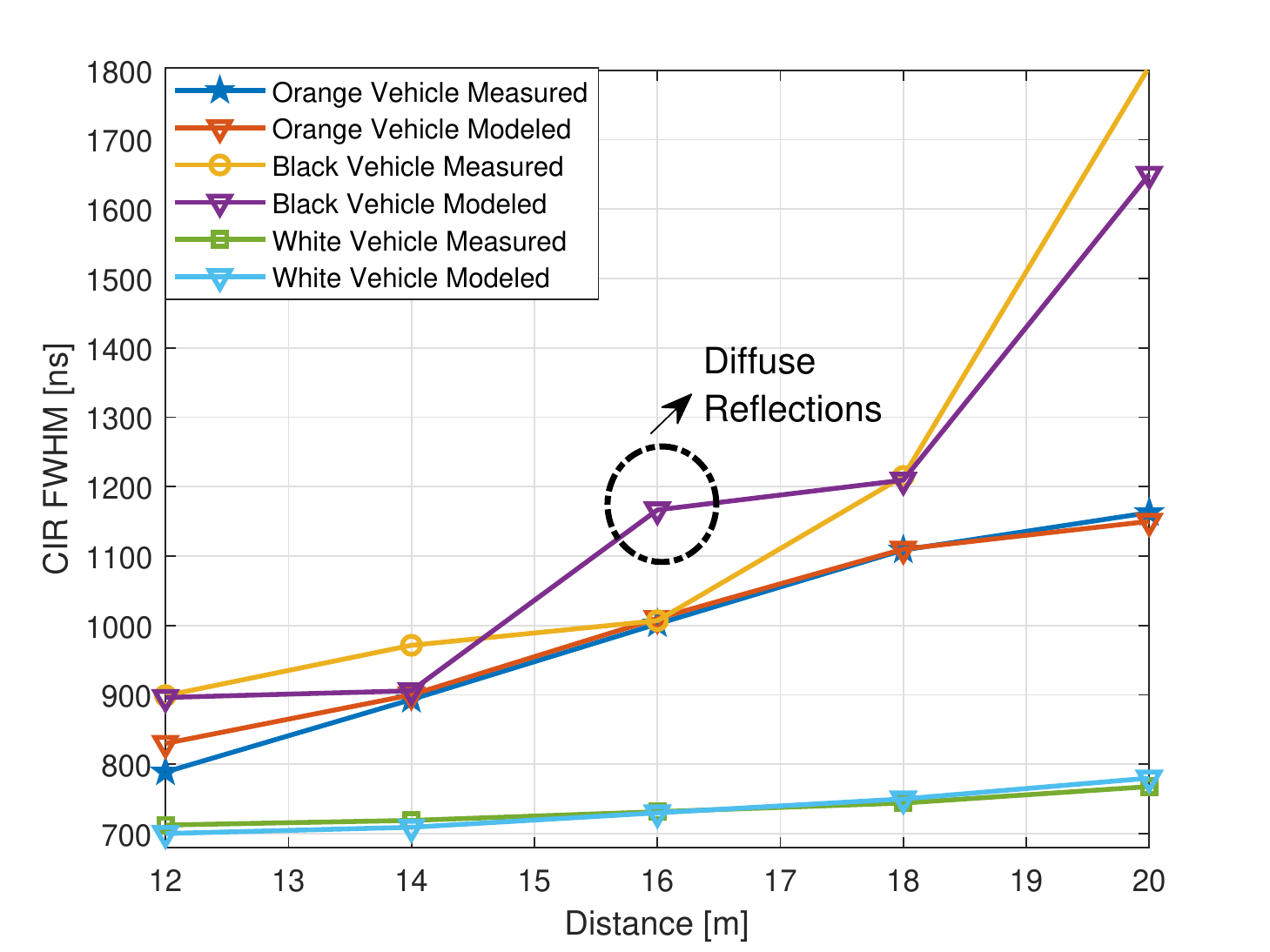} %Elongate.eps
\caption{CIR temporal broadening with increasing distances.}
\label{Fig:CIRBroadening}
\end{figure}

Fig.~\ref{Fig:CIRBroadening} shows the distance dependent measured and \ac{WDGF} modeled \ac{FWHM} of the \ac{LoS} and \ac{NLoS} channels  under consideration. For orange and white vehicle reflected \ac{NLoS} channels, \ac{WDGF} model well approximates the \ac{CIR} \ac{FWHM}, whereas for black vehicle reflected \ac{NLoS} channels, the model accuracy decreases for diffuse reflections, where model overestimates the measured \ac{CIR} \ac{FWHM} by 160 ns. At 16 m distance, black vehicle body diffuse reflections help to preserve \ac{CIR} \ac{FWHM} with 29 ns elongation, whereas the model estimates 261 ns broadening. However, providing 2 ns and 5 ns \ac{FWHM} estimation accuracy at 12 m and 18 m distances, respectively, for specular reflections of black vehicle induced \ac{NLoS} channel, the \ac{WDGF} model can be considered to approximate the \ac{NLoS} \ac{VVLC} \ac{CIR} for different distances and reflection surfaces. 

\subsection{Path Loss Model} The least-squares fitting based empirical \ac{NLoS} channel path loss model is obtained as 
\begin{align}
{\rm PL \; [dB]}= & 10{\rm log}_{10}\left(\left(\alpha {{\rm exp}(-n}\frac{ d_{0}}{d})\right)^{d-d_{0}} \left(\frac{d}{d_{0}}\right )^\beta \right) \nonumber \\
& + {\rm PL_{REF}}, \label{model3_coef}
\end{align}

\noindent where ${\rm PL_{REF}}$ is the path loss at the reference distance $d_{0}$, $d$ is the reflection surface to receiver distance, $\alpha$, $\beta$ and $n$ are the reflection surface dependent model coefficients. The proposed equation incorporates reflection surface dependent parameters into the path loss at reference distance, since each reflection surface inherits varying distance dependent reflection characteristics.

\begin{figure}[h]
\centering
\includegraphics[clip,trim=0 0 0 0,width=1\linewidth]{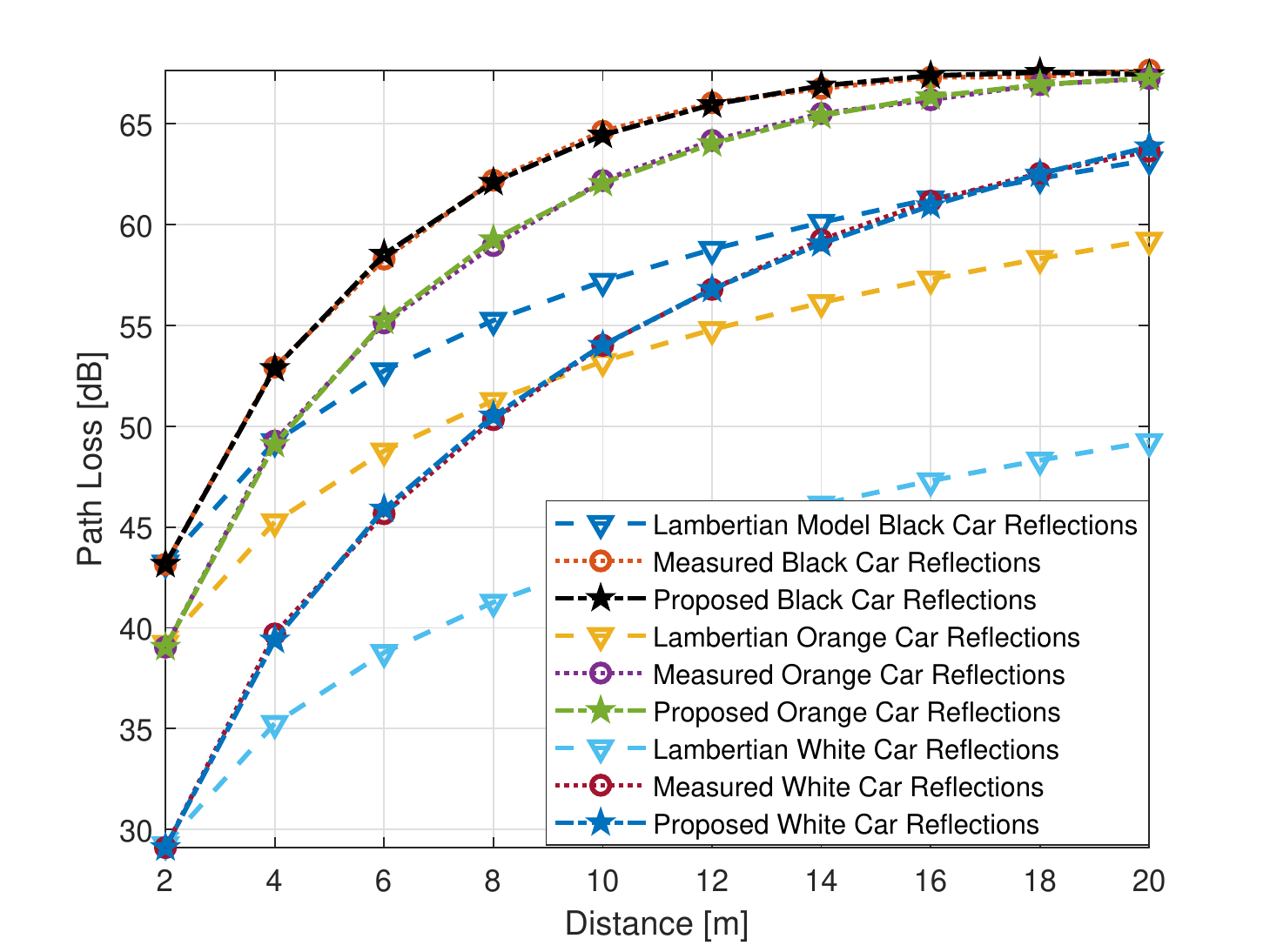}
\caption{\ac{VVLC} \ac{NLoS} path loss for measured, Lambertian single order reflection and proposed model fits}
\label{Fig:PLcomparison}
\end{figure}

Fig.~\ref{Fig:PLcomparison} shows measured path loss along with the single order reflection model detailed in Appendix, and proposed empirical \ac{NLoS} channel path loss models for three different vehicle reflection surfaces. The proposed path loss expression provides 0.18 dB, 0.19 dB, and 0.22 dB \ac{RMSE} to the measurement data. On the other hand, single surface reflection model underestimates the path loss by 14 dB, 8 dB, and 4 dB at 20 m reflector-receiver distance for white, orange and black cars, respectively. The proposed model, incorporating wide range of modulation frequencies of the \ac{LED}, yields accurate path loss estimations to evaluate the performance of multi-carrier based optical communication schemes such as \ac{DCO-OFDM}. 

\begin{figure}[h]
\centering
\includegraphics[clip,trim=0 0 0 0,width=\linewidth]{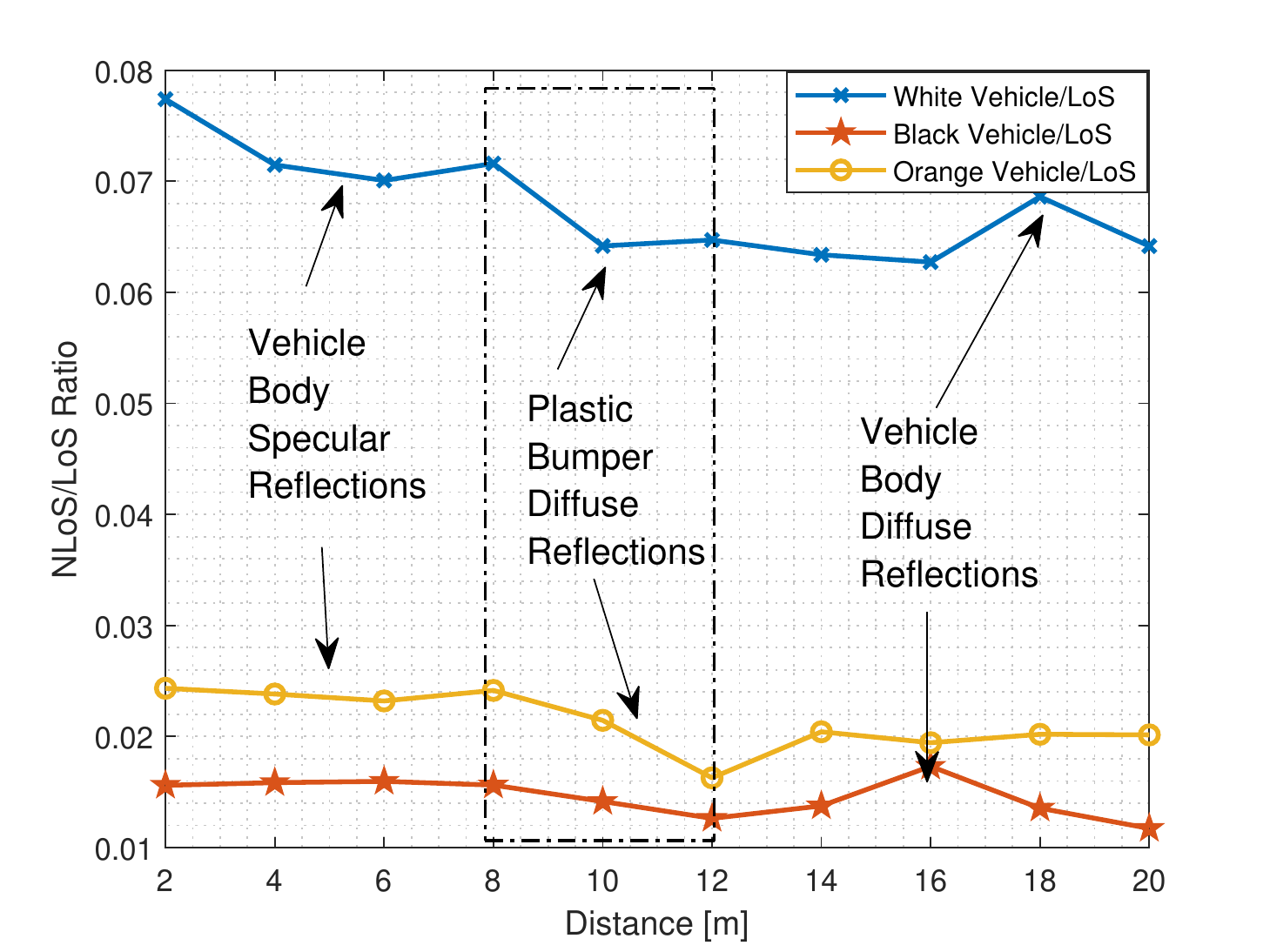}
  \caption{Ratio of NLoS to LoS channel gains for different reflective surfaces and distances.}
\label{Fig:Ratio}
\end{figure}

Fig.~\ref{Fig:Ratio} shows the normalized \ac{NLoS} channel gain for varying transmitter-receiver distances, to evaluate the effects of various surface and colors on \ac{NLoS} channel path loss. The ratio mainly depends on the reflection surface properties, where the receiver captures reflections from various parts of the vehicle surface, such as bumper, trunk lid and vehicle body, depending on the type of the vehicle. Considering the \ac{PD}'s \ac{FoV}, mainly specular reflections from vehicle body is captured up to 8 m. Then, reflections from bumpers, partially made up of rough plastic, yield diffuse reflections between $8-12$~m. Moreover, depending on the trunk lid design, flat surface of the orange vehicle yields specular reflections whereas the edges and curvatures of the black and white vehicle trunk lids lead to diffuse reflections between 12 m to 20 m.

\section{DCO-OFDM \ac{VVLC} Performance Analysis}
\label{sec:dcoofdm}

The aim of this section is to investigate the \ac{BER} performance of the \ac{DCO-OFDM} \ac{VVLC} scheme for \ac{NLoS} channel at varying distances with Monte Carlo simulations. Since \ac{VVLC} performance highly depends on the ambient noise levels, we investigate day and night performance of the \ac{DCO-OFDM} scheme with two different noise levels and modulation orders. The \ac{NLoS} \ac{VVLC} measured channel gain values are considered with the simulation parameters set including $P_{\rm T}=-12$~dBm, $N=64$, $R=0.3$, $M=4$-QAM for day, 16-QAM for night, noise power of $-100$~dBm for day and $-110$~dBm for night scenarios.      %summarized in Table~\ref{tbl:parameters}. 

\begin{figure}[h]
\centering
\includegraphics[clip,trim=0 0 0 0,width=\linewidth]{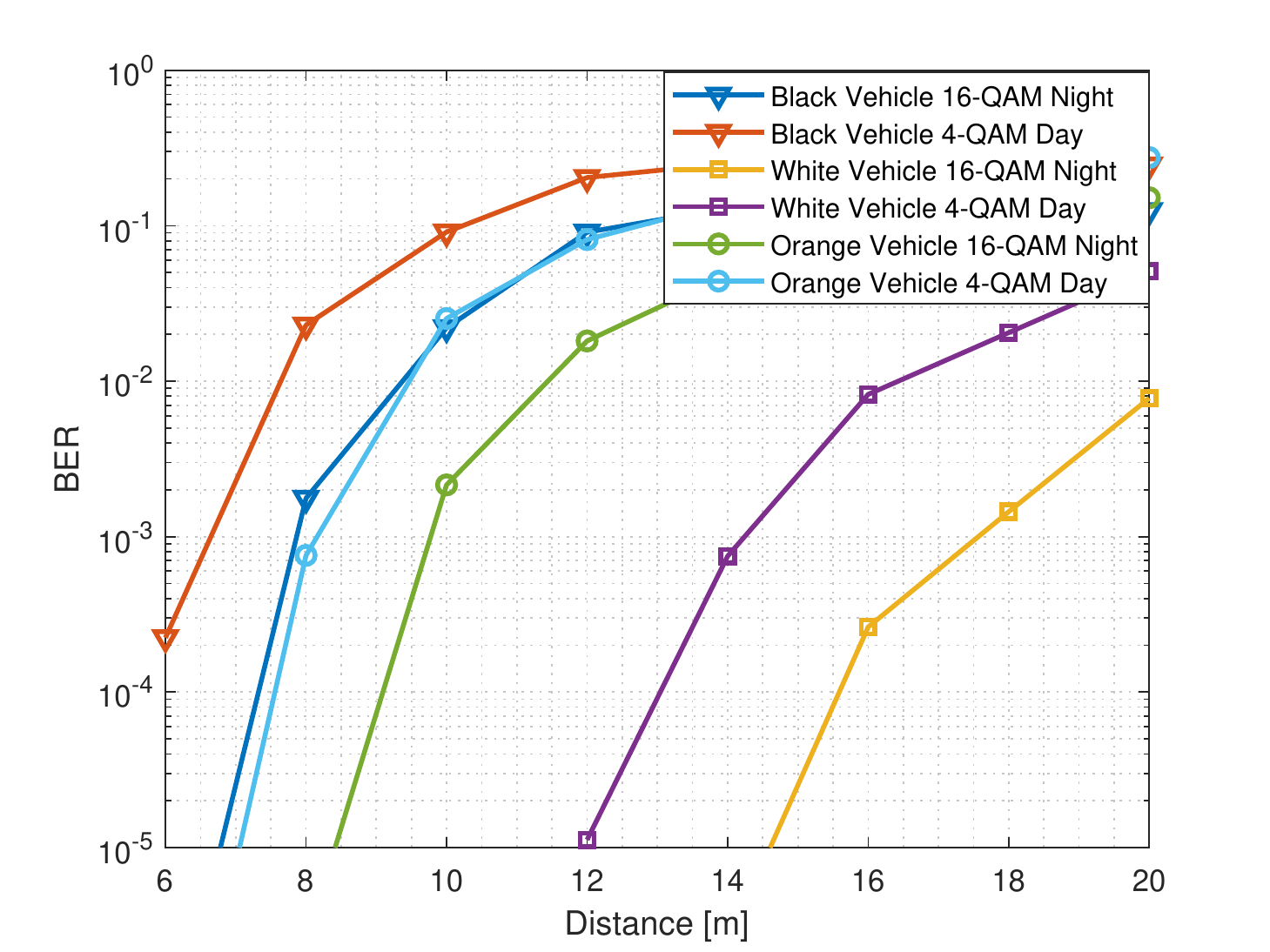}
\caption{\ac{NLoS} \ac{VVLC} DCO-OFDM \ac{BER} performance for three different vehicle surface reflections, day (4-QAM) and night (16-QAM) scenarios}
\label{Fig:NLOSBER}
\end{figure}

Fig.~\ref{Fig:NLOSBER} depicts the Monte Carlo simulation results for distance based \ac{BER} performance of the \ac{DCO-OFDM} scheme with the consideration of measured channel gain values. For black vehicle induced \ac{NLoS} channel, achievable reflector-receiver distances for target \ac{BER} of $10^{-3}$ are 6.67 m and 7.86 m for day and night conditions, respectively. Considering orange vehicle reflected \ac{NLoS} channel, the achievable distances for target \ac{BER} of $10^{-3}$ are 8.22 m and 9.79 m, whereas for white vehicle \ac{NLoS} channel, target \ac{BER} is attained at 14.26 m and 17.62 m distances at day and night conditions, respectively. The achievable distance performance of the white vehicle reflected \ac{NLoS} channel is substantially better than black and orange vehicle reflected channels, due to stronger reflections. Moreover, considering day and night performance of the \ac{NLoS} channels under consideration, the achievable distance difference between two scenarios (i.e., day 4-QAM and night 16-QAM) increases with stronger reflections, where the difference is 1.19 m, 1.57 m and 3.36 m for black, orange and white vehicle reflected \ac{NLoS} channels, respectively.

\section{Conclusion} \label{sec:conc}

In this paper, we investigate the \ac{NLoS} \ac{VVLC} channel path loss, \ac{CIR} and \ac{BER} performance for \ac{DCO-OFDM} based transmissions. \ac{NLoS} channel path loss expression is proposed by using the channel measurement data obtained through three different reflection surfaces, where it is demonstrated to well approximate the measurement data. Moreover, \ac{CIR} of the \ac{NLoS} \ac{VVLC} channel is demonstrated to broaden with the increasing distance due to scattering of photons. Therefore, \ac{WDGF} model is adopted to approximate the \ac{CIR} of \ac{NLoS} channel, where it provides higher accuracy for specular reflections induced \ac{NLoS} channels. The diffuse reflections due to vehicle tailgate design is demonstrated to yield channel gain and \ac{CIR} \ac{FWHM} fluctuations, where the approximation accuracy of the \ac{WDGF} \ac{CIR} model decreases. The \ac{BER} performance of \ac{DCO-OFDM} \ac{VVLC} scheme demonstrates that, white vehicle reflections yield substantially better achievable distance for target \ac{BER}, whereas \ac{NLoS} channel through the black and orange vehicle reflections are able to provide 6.64 m and 8.20 m link distance under day conditions, which can be practical for close proximity \ac{V2V} safety applications. The proposed path loss model along with the \ac{WDGF} \ac{CIR} models can be employed for system simulations.

\appendix[Lambertian Single Order Reflection Model]\label{SR}

In this appendix, widely used single order reflection model is provided to evaluate the path loss and time domain parameters of \ac{NLoS} \ac{VLC} channels. The single order reflection model is considered as a benchmark to evaluate the \ac{NLoS} \ac{VVLC} channel path loss. For single order reflection model, the $i^{\rm th}$ \ac{NLoS} component is obtained through the summation of the single order reflected components from each reflection surface.

The channel gain of the single order reflection \ac{NLoS} \ac{VLC} channel is given by~\cite{dixit2020performance}
\begin{align}
    H_{i}(d_{1},d_{2}) =  & \frac{(m+1)A_{\rm r}}{2\pi^2d_{\rm 1}^2d_{\rm 2}^2}\rho dA_{\rm r}f(\alpha, \beta, \Psi), \\
    f(\alpha, \beta, \Psi) = &  {\rm cos}^m\phi {\rm cos} (\alpha) {\rm cos} (\beta) T_{\rm s}(\Psi)g(\Psi) {\rm cos} (\Psi),
\end{align}    
% %\delta \times t-\left(\frac{d_{\rm j1}+d_{\rm j2}}{c}\right), \Psi<\Psi_{c},

\noindent where $A_{\rm r}$ is the active area of the \ac{PD}; $d_{\rm 1}$ is the distance between the transmitter and reflector vehicle; $d_{\rm 2}$ is the distance between the receiver and reflector vehicle; $\Psi$ is the angle of incidence at the receiver; $\Psi_{c}$ is the field of view (FOV); $\alpha$ and $\beta$ are the angle of irradiance to the reflector vehicle and incidence to the receiver, respectively; $\rho$ and $dA_{\rm r}$ are the reflection coefficient and small differential area of the reflector surface, respectively; $T_{\rm s}(\Psi)$ is the gain of an optical filter; $g(\Psi)$ is the gain of a non imaging optical concentrator with refractive index $n$ given by
\begin{equation}
    g(\Psi) = \frac{n^2}{sin^2(\Psi_{\rm c})}.
\end{equation} %, $c$  is the speed of light and $\delta(t)$ is the dirac delta function.

Single order reflection model generally assumes uniform distribution of the reflection points throughout the reflection area, which is applicable for large surface areas such as walls. Thus, single order reflection model is based on the assumption of perfect diffuse surface, which equally scatters incident illumination in all directions.%However, vehicle reflection surfaces incorporate different geometries due to varying design and functionality requirements.

%\section*{Acknowledgment}


\bibliographystyle{ieeetr}
\bibliography{references}

%\markboth{Submitted to IEEE Transactions on Vehicular Technology,~Vol.~XX, No.~XX, XXX~2015}
%{}
%{Shell \MakeLowercase{\textit{et al.}}: Bare Demo of IEEEtran.cls for Journals}

%\begin{IEEEbiography}{Yuguang ``Michael'' Fang}
%\end{IEEEbiography}

\end{document}